\documentclass[iop]{emulateapj}
\usepackage{graphicx}
\usepackage{subfigure}
\usepackage{rotating} 

\newcommand{\lsim }{{\lower0.8ex\hbox{$\buildrel <\over\sim$}}}
\newcommand{\gsim }{{\lower0.8ex\hbox{$\buildrel >\over\sim$}}}

\usepackage{soul}

\def\Chandra{\emph{Chandra}}
\def\Chandraacis{\emph{Chandra/ACIS}}
\def\RXTE{\emph{RXTE}}
\def\Swiftxrt{\emph{Swift/XRT}}
\def\Swiftbat{\emph{Swift/BAT}}

\def\MAXI{\emph{MAXI}}
\def\Maxigsc{\emph{MAXI/GSC}}

\def\simge{\mathrel{%
  \rlap{\raise 0.511ex \hbox{$>$}}{\lower 0.511ex \hbox{$\sim$}}}}
\def\simle{\mathrel{
  \rlap{\raise 0.511ex \hbox{$<$}}{\lower 0.511ex \hbox{$\sim$}}}}

\newcommand{\Msun}{\ifmmode {M_{\odot}}\else${M_{\odot}}$\fi}
\newcommand{\Lsun}{\ifmmode {L_{\odot}}\else${L_{\odot}}$\fi}
\newcommand{\Rsun}{\ifmmode {R_{\odot}}\else${R_{\odot}}$\fi}

\usepackage{xargs}
\usepackage{color}

\shorttitle{Discovery of Terzan 5 X-3}
\shortauthors{Bahramian et al.}

\begin{document}

\title{Discovery of the Third Transient X-ray Binary in the Galactic Globular Cluster Terzan 5 }

\author{Arash~Bahramian\altaffilmark{1},  Craig~O.~Heinke\altaffilmark{1}, Gregory R. Sivakoff\altaffilmark{1}, Diego Altamirano\altaffilmark{2,3}, Rudy Wijnands\altaffilmark{2}, Jeroen Homan\altaffilmark{4}, Manuel Linares\altaffilmark{5,6}, David Pooley\altaffilmark{7,8}, Nathalie Degenaar\altaffilmark{9,10}, Jeanette~C. Gladstone\altaffilmark{1}}

\altaffiltext{1}{Dept. of Physics, University of Alberta, CCIS 4-183, Edmonton, AB T6G 2E1, Canada; bahramia@ualberta.ca}
\altaffiltext{2}{Astronomical Institute 'Anton Pannekoek', University of Amsterdam, Science Park 904, 1098 XH Amsterdam, the Netherlands}
\altaffiltext{3}{Physics \& Astronomy, University of Southampton, Southampton, Hampshire SO17 1BJ, UK}
\altaffiltext{4}{Kavli Institute for Astrophysics \& Space Research, MIT, 70 Vassar St., Cambridge, MA 02139, USA}
\altaffiltext{5}{Instituto de Astrof{\'i}sica de Canarias, c/ V{\'i}a L{\'a}ctea s/n, E-38205 La Laguna, Tenerife, Spain}
\altaffiltext{6}{Universidad de La Laguna, Dept. Astrof{\'i}sica, E-38206 La Laguna, Tenerife, Spain}
\altaffiltext{7}{Dept. of Physics, Sam Houston State University, Huntsville, TX 77341, USA}
\altaffiltext{8}{Eureka Scientific, Inc., Oakland, CA, USA}
\altaffiltext{9}{Dept. of Astronomy, University of Michigan, 500 Church Street, Ann Arbor, MI 48109, USA}
\altaffiltext{10}{Hubble fellow}

\begin{abstract}
We {report} and study the outburst of a new transient X-ray binary (XRB) in Terzan 5, the third detected in this globular cluster, Swift J174805.3-244637 or Terzan 5 X-3.  We {find} clear spectral hardening in \Swiftxrt\ data during the outburst rise to the hard state, thanks to our early coverage (starting at $L_X\sim4\times10^{34}$ ergs/s) of the outburst.  This hardening appears to be due to the decline in relative strength of a soft thermal component from the surface of the neutron star (NS) during the rise. We identify a {Type I X-ray burst} in \Swiftxrt\ data with a long (16 s) decay time, indicative of {hydrogen burning on the surface of the} NS. We use \Swiftbat, \Maxigsc, \Chandraacis, and \Swiftxrt\ data to study the spectral changes during the outburst, identifying a clear hard-to-soft state transition.   We use a \Chandraacis\ observation during outburst to identify the transient's position.  Seven archival \Chandraacis\ observations show evidence for variations in Terzan 5 X-3's non{-}thermal component, but not the thermal component, during quiescence.  The inferred long-term time-averaged mass accretion rate, from the quiescent thermal luminosity, suggests that if this outburst is typical and only slow cooling processes are active in the neutron star core, such outbursts should recur every $\sim$10 years.
\end{abstract}


\maketitle

\section{Introduction}
{Transient low-mass X-ray binaries (LMXBs) experience long periods (often years to tens of years) of quiescence. In quiescence matter flowing from the companion builds up in the accretion disk, punctuated by outbursts when the accretion disk crosses a pressure and temperature threshold, increases in viscosity, and dumps large quantities of matter onto the accreting compact object \citep[e.g., see ][for a review]{Lasota01}.  Their outbursts go through phases of varying X-ray spectra. These phases are generally interpreted as indicating the changing relative contributions of Comptonized optically thin emission vs. blackbody-like emission from an accretion disk, as the accretion rate and geometry change \citep[e.g.,][for reviews]{Remillard06,Done07}.  These X-ray spectral states have been studied both for black hole LMXBs, and for NS LMXBs \citep[e.g., ][]{Hasinger89,Gierlinski02,Gilfanov03,Gladstone07,Lin07}, which show an additional component from the NS surface.

The spectra of NS LMXBs in quiescence ($L_X\simle 10^{33}$ ergs/s) include thermal radiation from the (usually hydrogen) NS atmosphere \citep[blackbody-like;][]{Zavlin96,Rajagopal96}, and often a harder nonthermal component, usually fit with a power-law \citep{Campana98a}.  Several NS LMXBs have shown rapid, strong variability in quiescence indicative of accretion events, which can sometimes be clearly attributed to variation in both the thermal and power-law components \citep[e.g.,][]{Rutledge02b,Campana04a,Cackett10a,Fridriksson11}.  The thermal component can be produced either by re-radiation of stored heat from the cooling NS \citep{Brown98}, or by low-level accretion \citep{Zampieri95,Deufel01}, which produce similar spectra \citep{Zampieri95}.  

 Studies of the spectra of black hole LMXBs during their decline from the low/hard state into quiescence (as $L_X$ falls below $10^{35}$ ergs/s) have found clear softening \citep{Corbel06,Corbel08,ArmasPadilla13b,Plotkin13}.  The softening of black hole LMXB spectra has been interpreted as a change in the origin of the X-ray emission, produced at low luminosities by either a radiatively inefficient hot flow \citep{Esin97,Gardner12} or synchrotron emission from a jet \citep{Yuan05,Pszota08}.  A similar softening in the spectrum from the accretion flow occurs in NS systems at similar luminosities, where emission from the NS surface can play a role \citep{ArmasPadilla11,Degenaar13b,Linares13}.

Thermonuclear X-ray bursts burn accumulated hydrogen and/or helium on the NS surface, producing blackbody-like emission with a rapid rise, cooling over timescales of seconds to minutes \citep{Lewin93,Galloway08}.  X-ray bursts occurring in hydrogen-poor environments (either due to no hydrogen being present in the accreted material, or hydrogen being stably burned during accretion) show different properties from those in hydrogen-rich environments. The ratio of energy released by fusion in a burst to energy released during accretion is lower for helium bursts compared to hydrogen bursts due to the lower energy available from fusion.

Helium bursts generally have faster rise and decline times, since hydrogen burning involves the CNO cycle and thus is limited by the speed of $\beta$-decays \citep{Fujimoto81}. Pure He bursts can be ignited in neutron stars that accrete hydrogen at low mass accretion rates (e.g., \citealt{Peng07}), but neutron stars known to be accreting hydrogen-poor material (ultracompact systems with white dwarf donors) never show evidence of hydrogen-rich bursts \citep{Galloway08}.  Some bursts from ultracompact systems are relatively long, but these ``giant'' (or ``intermediate-duration") bursts exhibit dramatic photospheric radius expansion, thought to be produced by a thick layer of accumulated He, which can accumulate only at low ($L_X<$0.01$L_{Edd}$) accretion rates  \citep{intZand05}. }

Globular clusters are highly efficient at producing X-ray binaries through dynamical interactions, such as the exchange of (heavy) NSs into pre-existing binary stars, replacing the lower-mass star in the binary.  Of perhaps 200 galactic LMXBs known to have reached {$L_X$}$\sim10^{36}$ ergs/s {}, 18 (including Terzan 5 X-3) are located in globular clusters, a factor of $\sim$100 overabundance per unit stellar mass compared to the galactic disk.  LMXBs are concentrated in the densest, most massive  globular clusters, which have the highest predicted rates of stellar interactions \citep[e.g.,][]{Verbunt87,Heinke03d}. 
Studying the number and types of LMXBs in different globular clusters can help us understand the dynamical processes that produce LMXBs in clusters.  
For example, identifying multiple LMXBs in one cluster has implications for interpreting observations of X-ray emission from extragalactic globular clusters (such as their luminosity functions), where multiple LMXBs cannot be resolved \citep[e.g.,][]{Sivakoff07}.  Before Terzan 5 X-3, no more than two bright LMXBs had been identified in any one globular cluster \citep{White01,Heinke10,Pooley10Atel}.  

Terzan 5 is a dense and massive globular cluster close to the center of our Galaxy \citep[d=5.9$\pm0.5$ kpc, ][]{Valenti07}, showing evidence of two separate stellar populations of different iron abundances, ages and helium content \citep{Ferraro09,D'Antona10}.  
Calculations of its stellar encounter rate suggest it may produce more X-ray binaries than any other Galactic globular cluster \citep{Verbunt87,Lanzoni10,Bahramian13}.
This status is supported by the largest population of known millisecond radio pulsars in any globular cluster, which are thought to be the descendants of LMXBs \citep{Ransom05,Hessels06}. Terzan 5 also hosts more than 50 known X-ray sources \citep{Heinke06b}, including a dozen likely quiescent {LMXBs} (again the most numerous in any cluster).

Outburst of transient LMXBs have frequently been observed from Terzan 5 \citep{Makishima81,Warwick88,Verbunt95}. \Chandra\ observed one such outburst in 2000 \citep{Heinke03b}, pinning down the location of an LMXB called EXO 1745-248\footnote{Note that the true identity of the transient seen by EXOSAT in the 1980s (leading to the EXO name) is not known.}, which was shown to have an unusually hard spectrum in quiescence during later \Chandra\ observations \citep{Wijnands05, Degenaar12}.  Another Terzan 5 outburst was identified in 2002 in \RXTE\ All-Sky Monitor data \citep{Wijnands02Atel}, but no imaging observations were taken. In 2010 an outburst was seen from an 11 Hz pulsar, IGR J17480-2446 (Terzan 5 X-2, \citealt{Bordas10Atel,Strohmayer10Atel1}), leading to a variety of detailed studies of the orbit and spin period, bursts, spectrum, burst oscillations, and evolution \citep[e.g.,][]{Papitto11,Chakraborty11,Miller11,Motta11,Cavecchi11,Linares11,Linares12,Patruno12,Papitto12,Altamirano12b}. A \Chandra\ observation identifying the outbursting source \citep{Pooley10Atel} allowed follow-up observations to track the crustal cooling of the NS \citep{Degenaar11a,Degenaar11b,Degenaar13}, while the even more precise moon occultation position \citep{Riggio12} permitted identification of the near-IR counterpart \citep{Testa12}. Another Terzan 5 outburst, in 2011, was identified as EXO 1745-248 through a \Chandra\ observation \citep{Pooley11Atel}, and showed a superburst (a very long and energetic X-ray burst, thought to be powered by the burning of a thick layer of carbon) at the beginning of the outburst \citep{Serino12,Altamirano12} 

In this paper we identify and study the outburst of the third transient X-ray binary in {the} globular cluster Terzan 5, {Swift J174805.3-244637 (henceforth Terzan 5 X-3)}. We detected this transient using \Swiftxrt  \citep{Wijnands12atel}, and identified spectral hardening in the rise of the outburst \citep{Heinke12atel},  a Type I X-ray burst \citep{Altamirano12atel}, and the quiescent X-ray counterpart in \Chandra\ images  \citep{Homan12atel}. In \S\ref{sec:data} we present {the} X-ray data used and describe our data extraction. In \S\ref{sec:result} we {derive the position of} Terzan 5 X-3 {by }comparing observations before and during the outburst, {analyze the spectral variation of the persistent emission throughout the outburst, study the properties of the thermonuclear burst, and analyze its quiescent X-ray spectrum}. Finally, we discuss {our results} in \S\ref{sec:disc}.

\section{Data Extraction}\label{sec:data}

\subsection{\Swiftxrt}\label{sec:siwftxrt}
We monitored Terzan 5 {up to a few times per week} for part of 2012 with the \Swiftxrt, covering the 0.3-10 keV energy range \citep{Burrows05}, as part of {our} monitoring campaigns of globular cluster X-ray transients \citep[see ][Altamirano et al. \emph{in prep.}]{Altamirano12}. This monitoring enabled us to observe the rising outburst of a new transient (and the 3rd known transient LMXB in this cluster) {first detected on July 6th, 2012} \citep{Wijnands12atel}. 

We used \Swiftxrt's  photon counting (PC) mode, which produces two{-}dimensional images, and windowed timing (WT) mode for which CCD data is collapsed into a one-dimensional image for fast readout. 
PC mode data {should be checked for} pile-up when the count rate exceeds 0.5 count s$^{-1}$. Pile-up is the recording of multiple photons as a single event, leading in the worst case to rejection of all events from the center of the point-spread function, or PSF. Our \Swiftxrt\ observations include 22 observations during the outburst, with 8 observations in WT and the rest in PC mode (Table \ref{swift_data}). 

We used HEASOFT 6.12 and FTOOLS\footnote{http://heasarc.gsfc.nasa.gov/ftools/} \citep{Blackburn95} to{ reduce and} analyze the data. We reprocessed the data with the FTOOLS \emph{xrtpipeline} and manually extracted data for spectral analysis. {We investigated every observation for pile-up, } following the \Swiftxrt\ pile-up thread\footnote{http://www.swift.ac.uk/analysis/xrt/pileup.php}, and extracted data from an annulus around the source in PC mode observations that suffered from pile-up.{We subtracted background from a circular region in the vicinity of the source in all PC observations.} The extraction region for WT data was chosen to be a box around the event array {(background subtraction was unnecessary for these countrates),} as discussed in the \Swiftxrt\ data reduction guide\footnote{http://heasarc.nasa.gov/docs/swift/analysis}. We extracted spectra in the 0.5-10 keV bandpass using FTOOLS \emph{xselect}, and created ancillary response function (ARF) files for each observations using FTOOLS \emph{xrtmkarf}. We performed spectral analysis using XSPEC 12.7.1 \citep{Arnaud96}. 

For heavily absorbed sources, WT data show low energy spectral residuals, which look like a ``bump'' in the spectrum, and cause spectral uncertainties in the $ \lesssim 1.0$ keV region\footnote{http://www.swift.ac.uk/analysis/xrt/digest\_cal.php}. We compared \Swiftxrt\  WT mode data to \Chandra\ data (\S \ref{sec:chandra}) taken within a few days, fitting them with the same model to find the energy range in which discrepancies appear. Based on this comparison, we ignored data below $1.4$ keV in all WT observations during the outburst.

\begin{table*}[h]
\begin{center}
\title{test?!}
\begin{tabular}{lcllll}
\hline
\hline
Obs. ID  		& 	Date 		&	MJD	&  	Exposure & 	avg. count rate 		& Notes\\
			&				&		&			&	(count s$^{-1}$)	&	\\
\hline
\multicolumn{2}{l}{\Swiftxrt\ observations} \\
\hline
32148002		&	2012-02-09	&	55966.9	&	985 s	&	1.75$\times10^{-2}$	&PC	mode; quiescent\\
91445001		&	2012-06-11	&	56089.8	&	913 s	&	2.67$\times10^{-2}$	&PC	mode; quiescent\\
91445002		&	2012-06-16	&	56094.5	&	1028	 s	&	1.48$\times10^{-2}$	&PC	mode; quiescent\\
91445003		&	2012-06-21	&	56099.0	&	1033	 s	&	1.66$\times10^{-2}$	&PC	mode; quiescent\\
91445004		&	2012-06-26	&	56104.8	&	1050	 s	&	9.68$\times10^{-3}$ 	&PC	mode; quiescent\\
91445005		&	2012-06-30	&	56108.6	&	935 s	&	1.40$\times10^{-2}$	&PC	mode; quiescent\\
91445006		&	2012-07-06	&	56114.8	&	1197	 s	&	7.65$\times10^{-2}$	&PC	mode; First detection of rise{;(1)}\\
32148003		&	2012-07-07	&	56115.8	&	987	s	&	0.154			&PC	mode{;(2)}\\
32148004		&	2012-07-08	&	56117.0	&	987	s	&	0.245			&PC	mode{;(2)}\\
32148005		&	2012-07-10	&	56118.1	&	781	s	&	1.26				&PC	mode; Piled up{;(2)}\\
32148006		&	2012-07-12	&	56120.7	&	978	s	&	3.19				&PC	mode; Piled up; Hard/soft transition\\
526511000	&	2012-07-13	&	56121.7	&	251	s	&	11.4				&WT	mode\\
91445008		&	2012-07-16	&	56124.3	&	253	s	&	14.2				&WT	mode\\
526892000	&	2012-07-16	&	56124.9	&	596	s	&	14.9				&WT	mode\\
32148007		&	2012-07-17	&	56125.9	&	960	s	&	20.4(16.3)$^a$		&WT	mode; Type I X-ray burst{;(3)}\\
91445009		&	2012-07-21	&	56129.1	&	1173	 s	&	71.8				&WT	mode\\
91445010		&	2012-07-26	&	56134.1	&	612	s	&	23.5				&PC	mode; Piled up\\
91445011		&	2012-08-01	&	56140.1	&	985	s	&	52.1				&WT mode\\
91445012		&	2012-08-05	&	56144.8	&	1057	 s	&	24.1				&WT	mode; Return to hard state?\\
91445013		&	2012-08-10	&	56149.4	&	1031	 s	&	4.15				&WT	mode\\
32148008		&	2012-08-11	&	56150.1	&	1488	 s	&	3.16				&PC	mode; Piled up\\
32148011		&	2012-08-13	&	56152.2	&	1480	 s	&	2.47				&PC	mode; Piled up\\
530808000	&	2012-08-13	&	56152.4	&	774 s	&	2.97				&PC	mode; Piled up\\
32148010		&	2012-08-14	&	56153.2	&	1494	 s	&	2.51				&PC	mode; Piled up\\
91445014		&	2012-08-15	&	56154.3	&	1060	 s	&	2.17				&PC	mode; Piled up\\
32148012		&	2012-08-19	&	56158.3	&	2153	 s	&	1.78				&PC	mode; Piled up\\
91445015		&	2012-08-20	&	56159.1	&	1006	 s	&	2.14				&PC	mode; Piled up\\
91445016		&	2012-08-24	&	56163.2	&	1131	 s	&	1.05				&PC	mode\\
32148013		&	2012-08-30	&	56169.3	&	2075 s	&	2.95$\times10^{-2}$	&PC	mode; quiescent\\
32148014		&	2012-09-13	&	56177.6	&	1953	 s	&	2.02$\times10^{-2}$	&PC	mode; quiescent\\
32148015		&	2012-09-20	&	56184.6	&	1976	 s	&	1.98$\times10^{-2}$	&PC	mode; quiescent\\
\hline
\multicolumn{2}{l}{\Chandraacis\ observations}\\
\hline
3798			&	2003-07-13	&	52833.6	&	39.34 ks	&	9.91$\times10^{-3}$	&	quiescent; {(4,5)}\\
10059		&	2009	-07-15	&	55027.7	&	36.26 ks	&	7.42$\times10^{-3}$	&	quiescent; {(6,7)}\\
13225		&	2011-02-17 	&	55609.3	&	29.67 ks	&	5.56$\times10^{-3}$	&	quiescent; {(6,7)}\\
13252		&	2011-04-29 	&	55680.7	&	39.54 ks	&	6.78$\times10^{-3}$	&	quiescent; {(7,8)}\\
13705		&	2011-09-05	&	55809.7	&	13.87 ks	&	5.69$\times10^{-3}$	&	quiescent; {(9)}\\
14339		&	2011-09-08	&	55812.1	&	34.06 ks	&	6.08$\times10^{-3}$	&	quiescent; {(9)}\\
13706		&	2012-05-13	&	56060.7	&	46.46 ks	&	7.58$\times10^{-3}$	&	quiescent; {(9)}\\
13708		&	2012-07-30	&	56138.4	&	9.84	ks	&	6.60	&	Terzan 5 X-3 outburst; Piled up; {(10,11)}\\
\hline
\end{tabular}
\end{center}
\caption{List of \Swiftxrt\ observations (top) and \Chandraacis\ observations (bottom) of Terzan 5 used. References: 1- \citealt{Wijnands12atel}, 2- \citealt{Heinke12atel}, 3- \citealt{Altamirano12atel}, 4- \citealt{Wijnands05}, 5- \citealt{Heinke06b}, 6- \citealt{Degenaar11a}, 7- \citealt{Degenaar11b}, 8- \citealt{Degenaar12}, 9- \citealt{Degenaar13}, 10- \citealt{Homan12atel}, 11- This work. a- The second count rate is calculated excluding the X-ray burst interval. Reported count rates are not corrected for pile up. MJDs are reported for start of each observation.}
\label{swift_data}
\label{chandra_data}
\end{table*}

\subsection{\Chandraacis}\label{sec:chandra}
We observed Terzan 5 X-3 during outburst with \Chandraacis\ in full-frame and FAINT telemetry mode with no grating (Obs. ID: 13708, PI: Pooley). We also used \Chandra\ archival data for our analysis of this source, details of which can be found in Table 1. All archival observations were taken in FAINT telemetry mode with the ACIS-S3 CCD at the focus. We analyze \Chandraacis\ data in the 0.5-10 keV energy range.

Data was reprocessed using CIAO 4.4 \citep{Fruscione06}, with CALDB 4.4.8, following the standard CIAO science threads\footnote{http://cxc.harvard.edu/ciao/threads/index.html}. We used observations during which all sources in the globular cluster were in quiescence.  We reprocessed the data, corrected the relative astrometry and ran CIAO \emph{reproject\_events}, and then stacked the event files together using CIAO \emph{dmmerge}.

Spectra were then extracted from both the archival and new data using CIAO task \emph{dmextract}. Terzan 5 X-3 was heavily piled up in the new observation (in outburst), and so we extracted a spectrum from the readout streak. Finally, we combined all archival (quiescent data) using FTOOLS \emph{addspec}\footnote{We found the results from CIAO's \emph{combine\_spectra} and FTOOLS's \emph{addspec} completely identical.}. Combining the quiescent data resulted in Å240 ks of exposure time. 

\subsection{MAXI/GSC}
\begin{figure*}[h]
\begin{center}
\includegraphics[scale=0.9]{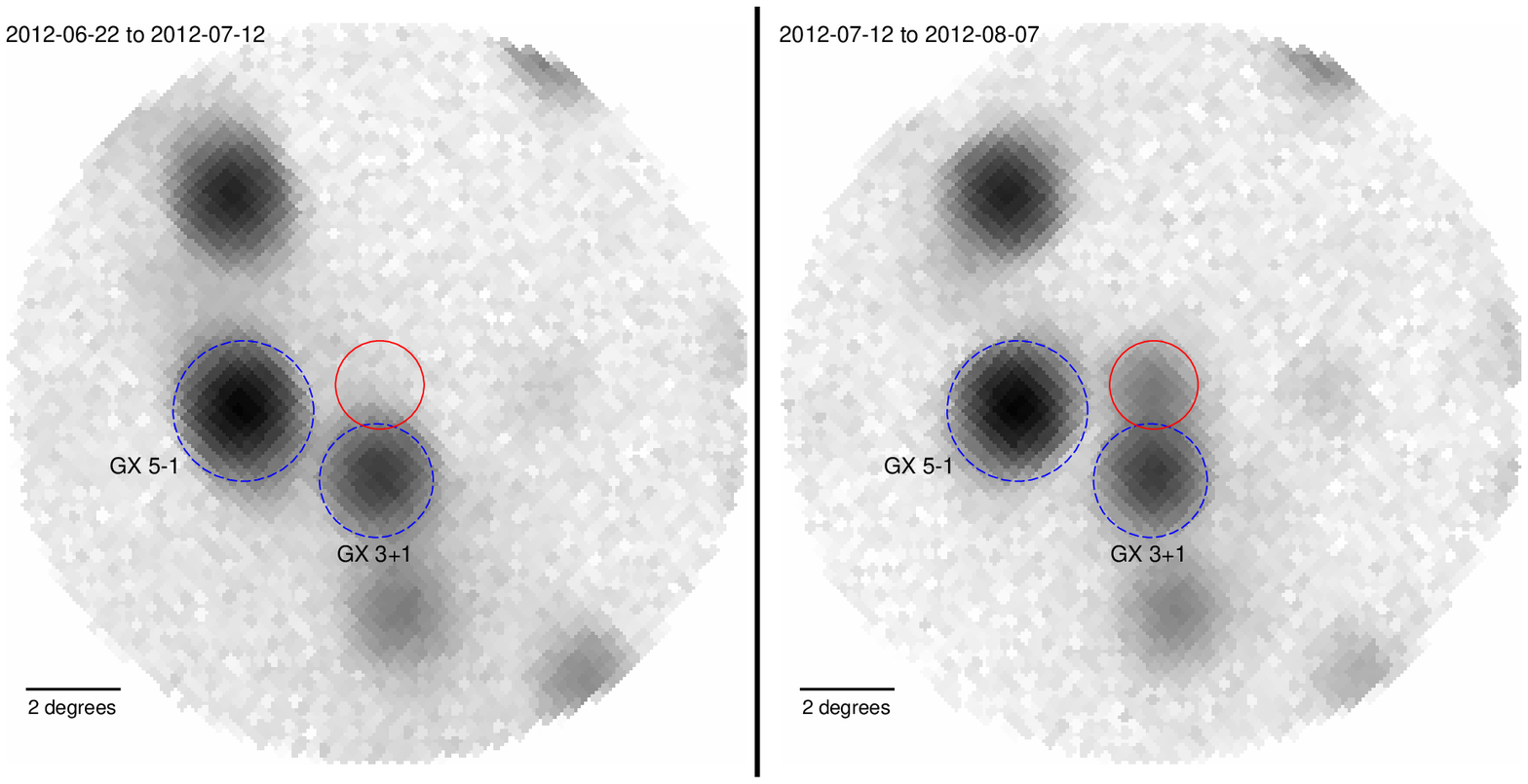} 
\caption{\Maxigsc\ 2-20 keV images of the sky around Terzan 5 (the red circle). Left: before Terzan 5 X-3 outburst. Right: During the outburst. \Maxigsc\ data for Terzan 5 may suffer  contamination from the nearby sources GX 5-1 and GX 3+1 (blue dashed circles). }
\label{maxi}
\end{center}
\end{figure*}

The \emph{MAXI} all sky X-ray monitor's  \citep{Matsuoka09} \emph{GSC} detector data covers the 2-20 keV energy range, and one-day averaged light curves are publicly provided in four bands: 2-4 keV, 4-10 keV, 10-20 keV and 2-20 keV \citep{Mihara11}. We noticed two problems with \Maxigsc\ light curves. Due to the low spatial resolution of \Maxigsc\ and bright sources in the crowded {field} of Terzan 5, there is the  possibility of background contamination from nearby sources like GX 3+1 and GX 5-1 (Fig.~\ref{maxi}). Since these two sources showed stable X-ray brightness with no obvious variations in the \Maxigsc\ data during Terzan 5 X-3 outburst, the background contamination may lead to a constant enhanced background.

We also noticed that \Maxigsc\ light curves show periodic behaviour with a period of $\approx$ 35 days for various well-known stable X-ray sources (e.g., the Crab {nebula}). This is caused by calibration issues regarding the 70-day precession of the International Space Station's orbit (\MAXI\ team 2013, \emph{priv.\ comm.}). This problem principally affects the 2-4 keV data, with less effect on the 4-10 and 10-20 keV light curves. Thus we ignored the 2-4 keV lightcurves for this research. We also ignored \Maxigsc\ 10-20 keV band lightcurves, due to the low statistical significance of Terzan 5 X-3's detection there.  We decontaminated the  \Maxigsc\ 4-10 keV lightcurve assuming a constant background count rate of 0.023, calculated based on a weighted average of the count rates before the outburst for a period of $\sim$120 days, with the addition of a systematic error based on the rms variations in the light curve before the outburst. We used the corrected values of the statistical uncertainties in \Maxigsc\ data, as the \MAXI\ team announced an erratum in the reported statistical uncertainties on April 26, 2013 (\MAXI\ team 2013, \emph{priv.\ comm.}) noting that the corrected uncertainties are a factor of 2 larger than previously reported.

\subsection{\Swiftbat}
The  \Swiftbat\ telescope data covers 15-150 keV energy range \citep{Barthelmy05}. We used daily light curves from the \Swiftbat\ transient monitor results provided by the \Swiftbat\ team \citep{Krimm13}. Data points on these daily light curves are from \Swiftbat\ survey data and are represented in a single band (15-50 keV). These points are the weighted average of all observations performed each day. The Swift/BAT has better angular resolution (20Õ vs.1 degree FWHM for Swift/BAT vs. MAXI/GSC). Such an improvement in the angular resolution limits the chances of contamination occurring. Although we cannot rule out some contamination, it seems reasonable that it does not pose a serious problem. As such, it is not surprising that we do not see evidence of contamination of the Swift/BAT data by nearby sources as is identifiable in the MAXI data.

\section{Analysis and Results}\label{sec:result}
\subsection{Position}\label{sec:position}
\begin{figure*}[h]
\begin{center}
\includegraphics[scale=0.9]{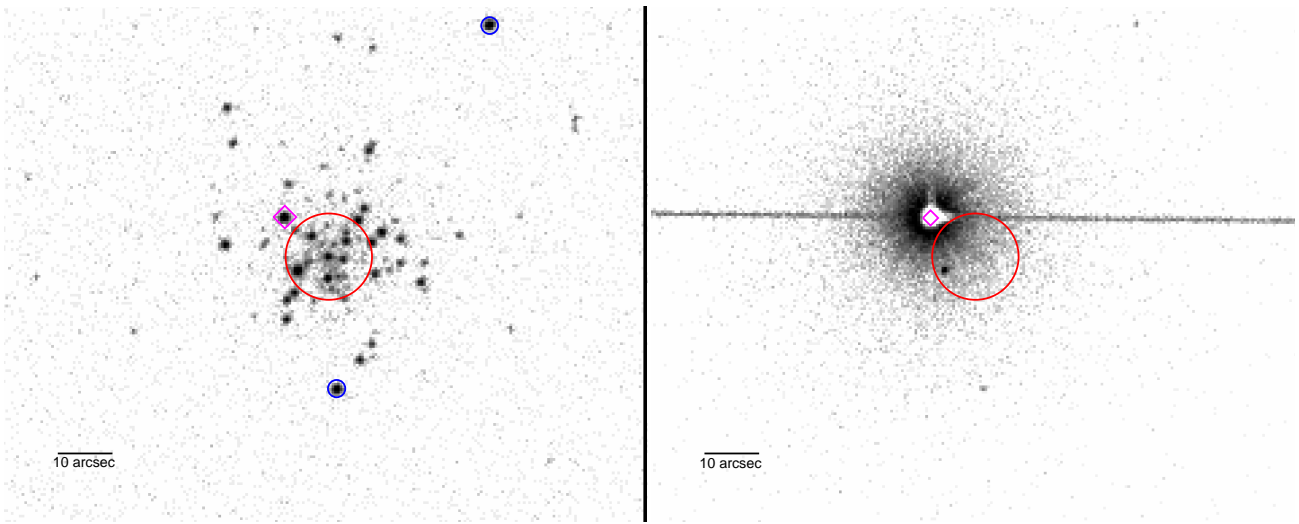}
\caption{\Chandraacis\ observations of Terzan 5. Left: Stacked image of observations during quiescence {(Bottom of Table \ref{chandra_data})} of all sources; {total exposure $\approx$ 240 ks}. Right: Outburst of Terzan 5 X-3 {(Obs.~ID 13708 with exposure $\approx$ 10 ks)}. The red circles and magenta diamond represent the core of Terzan 5{ \citep{Harris96}} and the position of Terzan 5 X-3, respectively (before  astrometric corrections to the outburst image). Blue circles identify additional sources (CXOGLB J174804.7-244709 and CXOGLB J174802.6-244602) used for constraining the hydrogen column density of Terzan 5 (\S\ref{nh}). }
\label{chandra}
\end{center}
\end{figure*}

We accurately and precisely located the position of Terzan 5 X-3, using the \Chandraacis\ data. We compared the \Chandraacis\ observation taken during the outburst and a stacked image of 7 \Chandraacis\ observations taken when all Terzan 5 sources were in quiescence (\S \ref{sec:chandra}, Table \ref{chandra_data}). We corrected the astrometry in the outburst observation by comparing the positions (using the CIAO \emph{wavdetect} tool) of three other sources in this observation with their astrometrically corrected positions as reported in \citet{Heinke06b}. 
Using a weighted average of the required shifts (+0.23$''$ for RA and +0.04$''$ for Dec), we find the position of Terzan 5 X-3 to be RA= 17:48:05.41$ \pm 0.02^s$ and Dec=$-$24:46:38.0$\pm0.2''$, in agreement (2$\sigma$) with the published position of the X-ray source CXOGLB J174805.4-244637 (\citealt{Heinke06b}), at RA= 17:48:05.413$\pm0.001 ^s$ and Dec=$-$24:46:37.67$\pm0.02''$ (Fig.~\ref{chandra}). 

\subsection{Phases of the outburst}\label{phases}
\begin{figure*}[h]
\begin{center}
\includegraphics[scale=0.42]{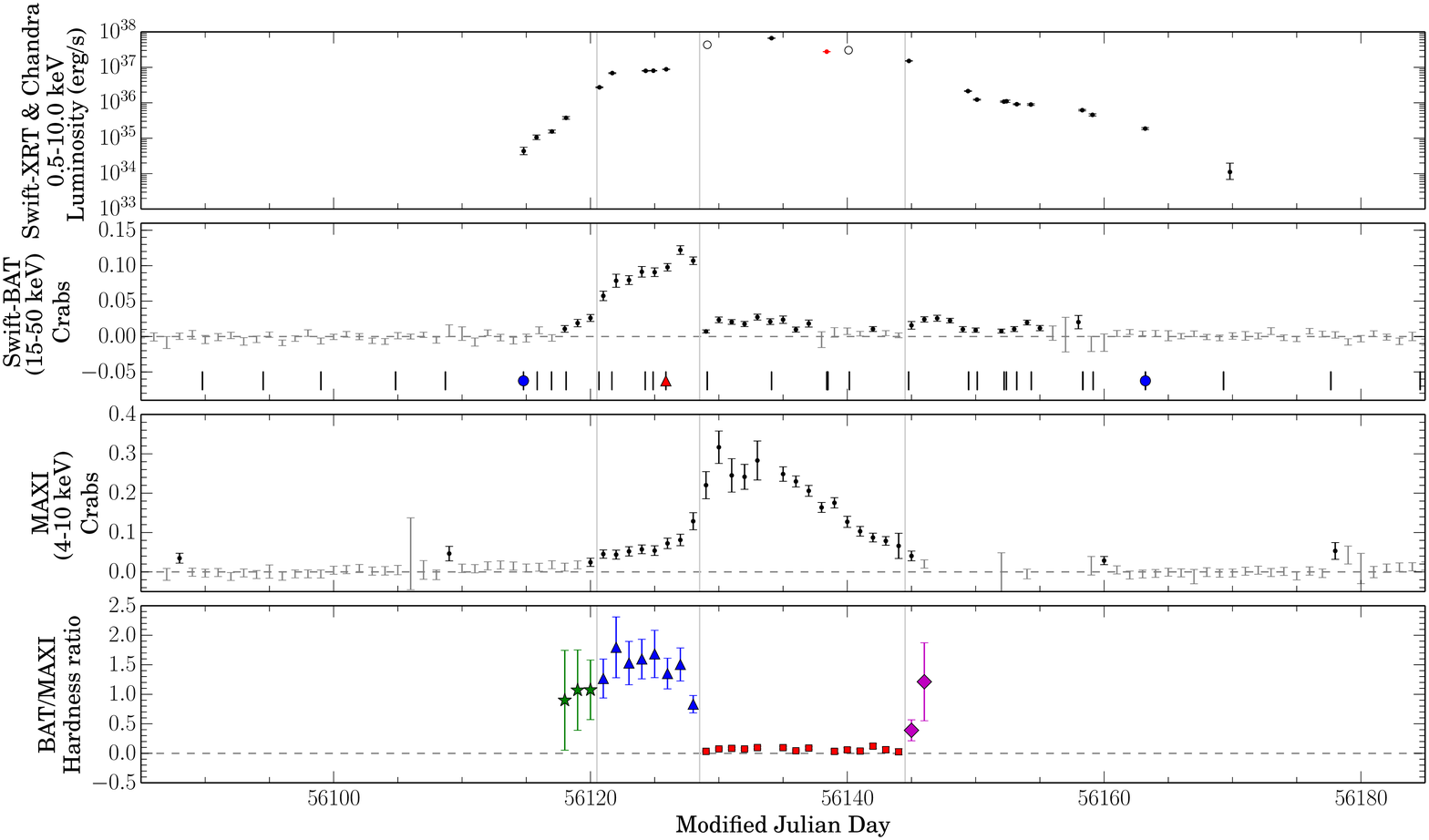}
\caption{Evolution of Terzan 5 X-3 outburst. From the top, First panel: Luminosities from spectral fitting of \Swiftxrt \& \Chandraacis\ pointed observations (\S \ref{outburst}). Empty circles represent observations where these fits were poor ($\chi_\nu^2 > 2$). The red datapoint represents \Chandraacis\ observation. Second panel: \Swiftbat\ lightcurve in the 15-50 keV band. Third panel: \Maxigsc\ background-subtracted lightcurve in the 4-10 keV band. Fourth panel: Hardness ratio (H/S after conversion to Crab units) of the two light curves (\Swiftbat/\Maxigsc). Black data points in the upper two panels represent significant detections ($> 2\sigma$) in each band, while grey bars show times when Terzan 5 X-3 was not significantly detected.  Vertical lines at the bottom of the first panel represent times of pointed \Swiftxrt\ and \Chandraacis\ observations. The first and last detection of the outburst in \Swiftxrt\ data is represented by blue circles. The red triangle identifies the time of the detected thermonuclear burst (\S \ref{thermo}). Colors and shapes in the {bottom} panel indicate the different phases of outburst; a) green stars, rise, b) blue triangles, hard state, c) red squares, soft state, d) magenta diamonds, decline. Both \Swiftbat\ and \Maxigsc\ lightcurves are in Crab units{ and error bars are 1$\sigma$ uncertainties}. Vertical lines in panels show approximate bounry of introduced phases. Note that in the fourth panel symbol size is larger than the errorbars in the soft state. All daily averages are plotted at the beginning of each day.}
\label{lc}
\end{center}
\end{figure*}

We used \Maxigsc\ and \Swiftbat\ hard X-ray transient monitor light curves to study the evolution of Terzan 5 X-3's outburst in the soft X-ray (4-10 keV, \Maxigsc) and hard X-ray (15-50 keV, \Swiftbat) bands. We converted count rates into equivalent fluxes from the Crab Nebula to make these light curves suitable for comparison. For this purpose, we used conversion coefficients given for each instrument\footnote{\MAXI - http://maxi.riken.jp/top/index.php?cid=000000000036 \Swiftbat - http://swift.gsfc.nasa.gov/docs/swift/results/transients}: for the \Swiftbat\ hard X-ray transient monitor 1 Crab = 0.22 count cm$^{-2}$ sec$^{-1}$ and for the \Maxigsc\ 4-10 keV band 1 Crab = 1.24 count cm$^{-2}$ sec$^{-1}$. We plot the light curves for Terzan 5 X-3's outburst as seen by both \Maxigsc\ and \Swiftbat, and their ratio, in Fig.~\ref{lc}.

We distinguish four phases of Terzan 5 X-3's outburst. {(a) }Rise: the hard X-ray brightness of the source increases, the source eventually becoming significantly detected in the soft X-ray as well. {(b) }Hard state: hard X-ray brightness reaches its peak. {(c) }Soft state: soft X-ray brightness peaks while hard X-ray brightness drops. {(d) }Decline: the source briefly gets brighter in the hard X-ray again before turning off. Unfortunately there is insufficient data from the MAXI /GSC during the decline of the outburst. Therefore, we are unable to confirm that Terzan 5 X-3 returns to the hard state during its decline. 

\begin{figure*}[h]
\begin{center}
\includegraphics[scale=0.39]{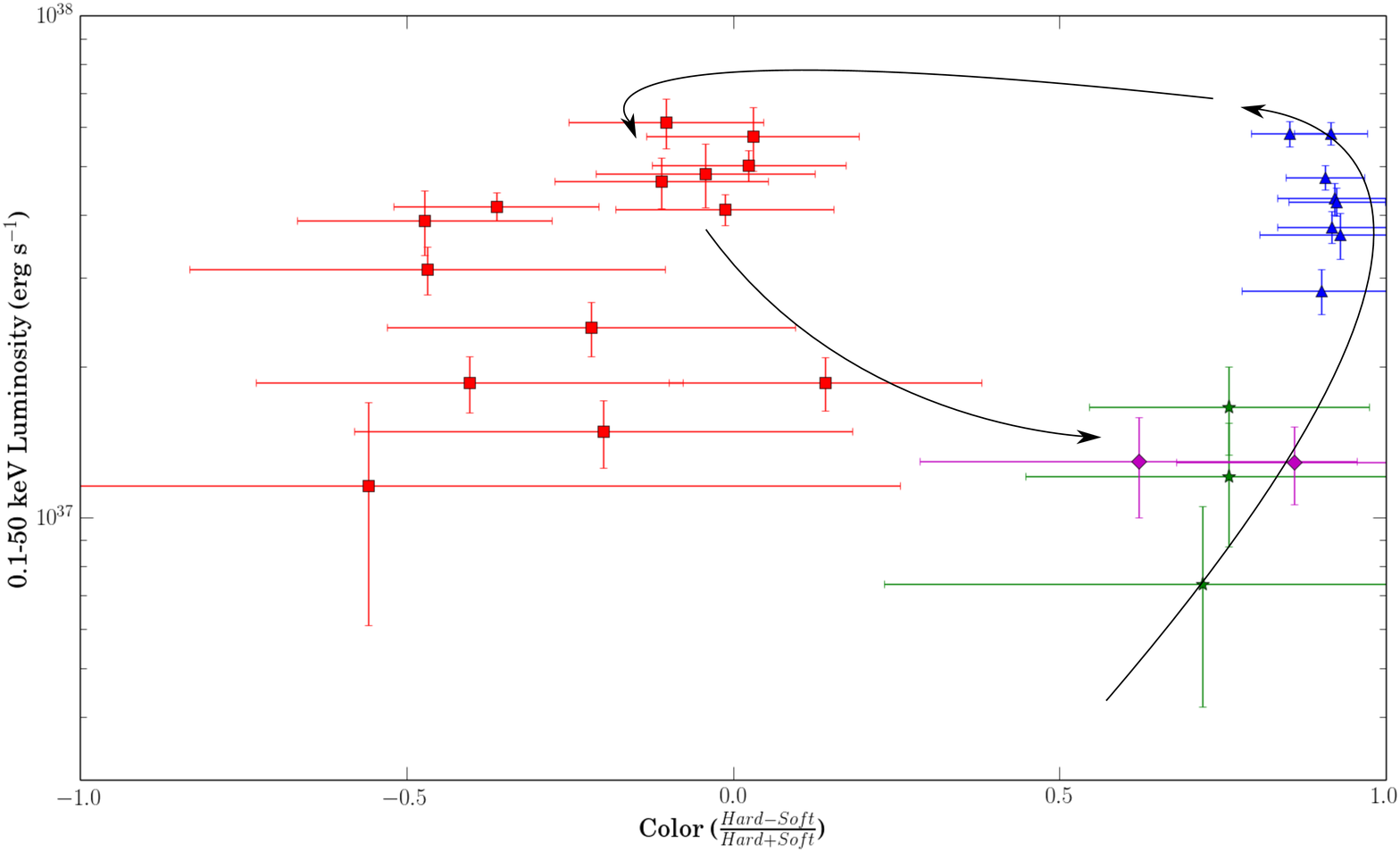}
\caption{Color-{luminosity} diagram of Terzan 5 X-3 outburst, using the soft X-ray (\Maxigsc\ 4-10 keV, S) and hard X-ray (\Swiftbat\ 15-50 keV, H) light curves, each in ergs s$^{-1}$ (\S \ref{phases}). color is (H-S)/(H+S), while luminosity is (H+S). colors and shapes of points indicate the phases of outburst (as in Fig.~\ref{lc}); a) green stars, rise, b) blue triangles, hard state, c) red squares, soft state, d) magenta diamonds, decline. Arrows represent the chronological order of data points. Error bars indicate 1$\sigma$ uncertainties.}
\label{ci}
\end{center}
\end{figure*}

We used the soft X-ray (\Maxigsc\ 4-10 keV, S) and hard X-ray (\Swiftbat\ 15-50 keV, H) lightcurves, to create a color-{luminosity} diagram (as an analogy to a hardness-intensity diagram, \citealt{Fender04}) for the outburst (Fig.~\ref{ci}). We defined our color to be (H-S)/(H+S), and defined {luminosity} as (H+S), converting count rates to luminosities in each band before their summation or subtraction. We converted count rates to luminosities with the assumption of a 5.9 kpc distance and power-law spectra, using power-law index values inferred from \Swiftxrt\ data spectral fitting (\S \ref{outburst}). We extrapolated the \Maxigsc\ 4-10 keV band flux to 0.1-12 keV and the \Swiftbat\ 15-50 keV band flux to 12-50 keV, and calculated luminosities in the 0.1-50 keV band. We cannot measure the spectral index above 10 keV, but since in the hard state the spectra are typically reasonably described by a power-law up to 50 keV, and in the soft state the flux above 12 keV is a minor contribution to the total, this is unlikely to have a large effect. In the 0.5-10 keV band, the source was bright for approximately 20 days, reaching a maximum luminosity of 7$\times$10$^{37}$ erg s$^{-1}$ and an average luminosity of 3$\times$10$^{37}$ erg s$^{-1}$ in this time interval.
The evolution of the outburst and phases mentioned above can be clearly seen in this color-luminosity diagram (Fig.~\ref{ci}). {Spectral evolution during the outburst, including at fainter fluxes but with a more limited bandpass, can also be seen in \Swiftxrt\ observations (\S\ref{outburst}).}

\subsection{Spectral analysis}
\subsubsection{Hydrogen column density of Terzan 5}\label{nh}
The first step in our spectral analysis is constraining hydrogen column density $N_H$,  which we check by spectral analysis of multiple sources in the cluster. Since the sources are located within 1' of each other, we expect little variation in $N_H$ along the different sightlines.  Except for LMXBs observed at high inclination, generally the measured $N_H$ throughout an outburst appears to be stable \citep{Miller09}, so we assume that Terzan 5 X-3's $N_H$ remains constant. 
We used \Chandraacis\ observations of Terzan 5 taken when all sources were quiescent (Table \ref{chandra_data}). We extracted spectra of three of the brighter faint sources in Terzan 5 (Terzan 5 X-3, CXOGLB J174804.7-244709 and CXOGLB J174802.6-244602; \citealt{Heinke06b}{, Fig.~\ref{chandra}}) from each observation using CIAO \emph{dmextract} and combined the extracted spectra for each source. We fit these combined spectra, along with the \Chandra\ spectrum of Terzan 5 X-3 during the outburst, simultaneously. We used appropriate models based on previous studies of each faint source \citep{Heinke06b}, confirmed as acceptable fits.  For Terzan 5 X-3 during quiescence, and for CXOGLB J 174804.7-244709, we used a NS atmosphere (NSATMOS, \citealt{Heinke06a}) plus a power-law (PEGPWRLW).  For CXOGLB J 174802.6-244602 we used a power-law. Finally, for Terzan 5 X-3 during outburst, we used a disk model (DISKBB) plus a thermal Comptonization model (COMPTT, \citealt{Titarchuk94}). We fit all these spectra simultaneously with a single value of $N_H$, using the PHABS model in XSPEC, with \citet{Anders89} abundances, finding $N_H = 1.74_{-0.08}^{+0.06}\times10^{22}${ cm$^{-2}$}. Individual spectral fits gave consistent results, with differences below the 10$\%$ level.
This value is consistent with the measurements in \citet{Heinke03b}, \citet{Wijnands05}, \citet{Heinke06b}, and \citet{Degenaar11a}, though not with the lower value of \citet{Miller11}\footnote{\citet{Miller11} report N$_H$=(1.17$\pm0.04)\times$10$^{22}$ cm$^{-2}$, though they do not report the abundance scale they use.}, and is consistent with the {\it E(B-V)} estimates in \citet{Valenti07} and \citet{Massari12}, using the \citet{Guver09} relation to $N_H$. We note that \citet{Miller11} fit a simple blackbody plus power-law to the outburst spectrum of IGR J17480-2446 (Terzan 5-X2), which may lead to systematic differences compared to more complex outburst spectral models. 
 We used our best-fit $N_H$ value in the PHABS model, using default abundances, for the rest of the spectral analysis throughout this paper.  
 
 For comparison, we also used the same procedure for constraining $N_H$ using the TBABS model (instead of PHABS), with \citet{Wilms00} abundances, finding an increase in $N_H$ to $2.6\pm0.1\times10^{22}$ cm$^{-2}$.  The remaining parameters agreed within the errors with the results using our default absorption model, indicating that for our level of analysis only internal consistency is required in the choice of absorption model.

\subsubsection{Outburst}\label{outburst}
The spatial resolution of \Swiftxrt\ is such that we must account for possible contamination due to emission form other XRBs in the cluster. To assess these levels, we fit the spectra of pre-outburst observations (using the same extraction region as for the outburst spectra), finding a background level of $L_X\sim10^{34}$ ergs/s. We used 6 \Swiftxrt\ observations taken before the outburst to estimate the combined spectrum from the other cluster sources. Due to the low number of counts per observation, we combined their spectra. We fit the resulting pre-outburst spectrum with an absorbed power-law model (Table \ref{pre-outburst-table}) with $N_H$ fixed to $1.74\times10^{22}${ cm$^{-2}$}, based on the results in \S \ref{nh}.  

\begin{figure*}[h]
\begin{center}
\includegraphics[scale=0.39]{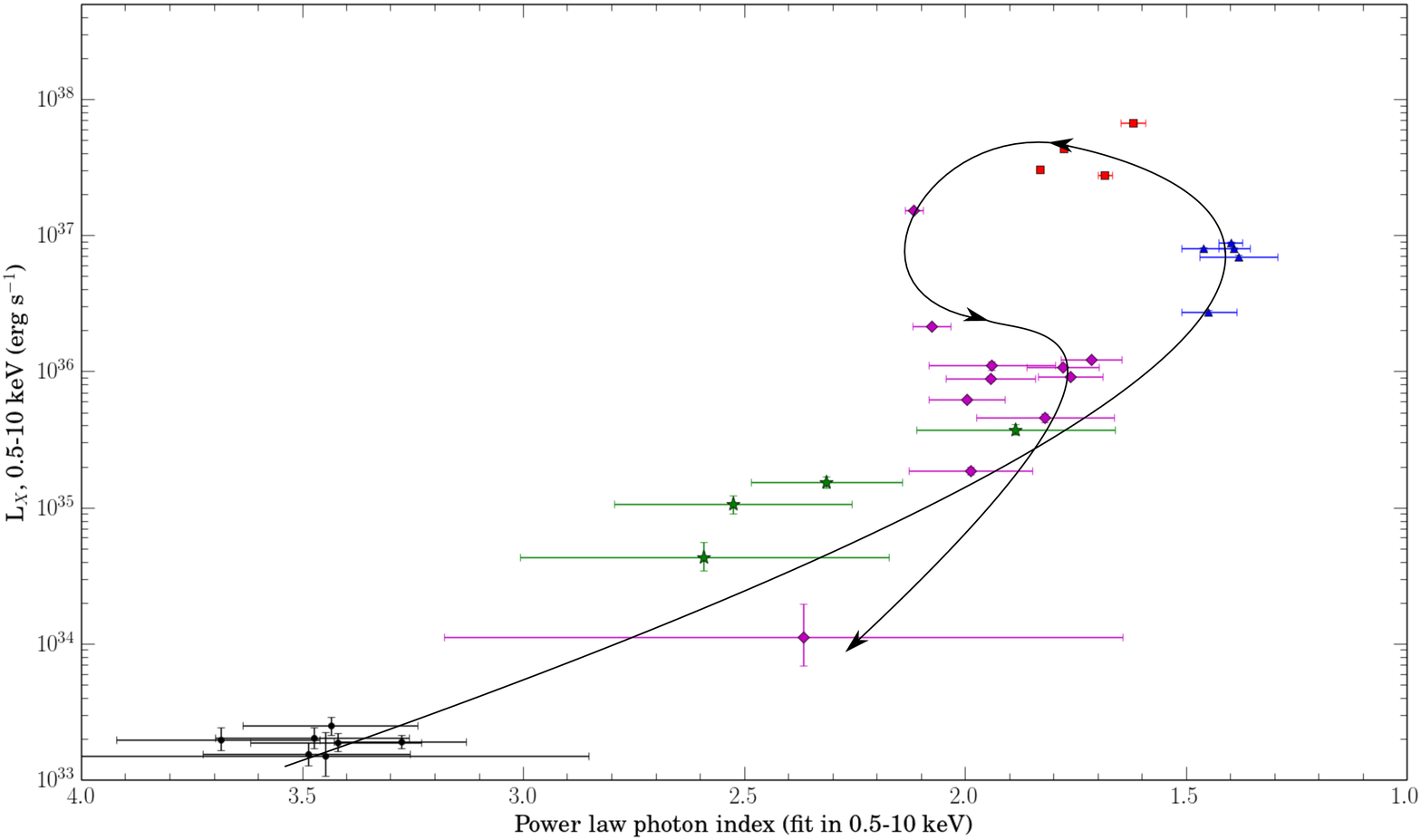}
\caption{Luminosity versus power-law photon index of Terzan5 - X3 during the outburst, using \Swiftxrt\ and one \Chandraacis\ outburst observations, plus two quiescent \Chandraacis\ observations. colors and shapes of points indicate the phases of outburst (as in Fig.~\ref{lc}); a) green stars, rise, b) blue triangles, hard state, c) red squares, soft state, d) magenta diamonds, return to hard state. The black circles represent \Chandraacis\ observations in quiescence, covering the range of observed hardness in the quiescent state. Arrows represent the chronological order of data points. Error bars indicate 1-$\sigma$ uncertainties.}
\label{spec_ev}
\end{center}
\end{figure*}

For our initial spectral analysis, we used an absorbed power-law model to fit Terzan 5 X-3's spectrum, including a second power-law component with values fixed to the pre-outburst results to model the background (Table \ref{powerlaw_data}, Fig.~\ref{spec_ev}). A simple power-law model provided a good fit to most of the spectra (Table~\ref{powerlaw_data}), while physically motivated complex spectral models could not be well-constrained for the high-quality bright outburst spectra (see below), so we focus on the results from power-law fits. This spectral analysis shows a significant drop of photon index {from 2.6$\pm$0.7 to 1.4$\pm$0.1 }during the outburst rise (Fig.~\ref{spec_ev}), showing a clear hardening of the spectrum during the rise from quiescence. During the hard state, the photon index shows no significant variations and is $\approx 1.4$. After the phase transition to the soft state, the photon index softens to $\approx 1.9$, with significant variations and several poor fits.

Several spectral models have been suggested in the literature to model the detailed spectra of transient NS LMXBs{ in outburst}, see e.g., \citet{Lin07}. These models usually contain a soft component for the radiation from the disk and/or boundary layer (i.e., a multi-color black body) and a hard component for the radiation from the hot corona around the accreting object (i.e., Comptonized radiation). We attempted to perform analyses of Terzan 5 X-3's outburst spectra using a variety of complex  models with multiple components (e.g., DISKBB + COMPTT, BBODY+COMPTT, BBODY+BKNPOW). {We could not} obtain strong constraints on the spectral parameters due to the limited energy band available and limited statistical quality of the XRT data.  
 The \Swiftxrt\ WT spectra suffered particularly from calibration issues below $\sim$1 keV(\S\ref{sec:siwftxrt}). The \Swiftbat\ survey mode hard ($>$15 keV) X-ray spectra had a low signal to noise ratio, which prohibited spectral analysis. We defer further detailed spectral fitting, e.g. to clearly distinguish thermal vs. nonthermal components, to future work.

\begin{table*}[h]
\begin{center}
\begin{tabular}{cccccccccc}
\hline
\hline
Obs. ID 		& MJD 		& Photon index 	& 	Flux  				&	L$_X$			&  $\chi_\nu^2$/D.O.F(nhp)\\
\hline
32148002-91445005 & 55966-56108 &2.4$_{-0.4}^{+0.5}$ & 	2.4$_{-0.6}^{+0.7}$	&	1.0$_{-0.2}^{+0.3}$	&	0.68/7(0.69)\\
\hline
91445006		&	56114.8	&	2.6$\pm$0.7	&	10$_{-3}^{+5}$			&	4$_{-1}^{+2}$		&	0.83/6(0.54) \\
32148003		&	56115.8	&	2.5$\pm$0.4	&	25$_{-5}^{+7}$			&	10$_{-2}^{+3}$		&	0.85/5(0.51) \\
32148004		&	56117.0	&	2.3$\pm$0.3	&	37$\pm6$				&	15$\pm$2			&	1.86/9(0.052) \\
32148005		&	56118.1	&	1.9$\pm$0.4	&	90$\pm12$			&	37$\pm$5			&	0.54/6(0.78) \\
32148006		&	56120.7	&	1.4$\pm$0.1	&	658$\pm32$			&	273$\pm$13		&	1.58/19(0.05) \\
526511000	&	56121.7	&	1.4$\pm$0.1	&	1658$\pm91$			&	688$\pm$	38		&	1.20/14(0.27) \\
91445008		&	56124.3	&	1.46$\pm$0.08	&	1920$\pm63$			&	797$\pm$	26		&	0.69/25(0.87) \\
526892000	&	56124.9	&	1.39$\pm$0.06	&	1936$\pm46$			&	803$\pm$	19		&	1.56/34(0.03) \\
32148007		&	56125.9	&	1.40$\pm$0.04	&	2125$\pm40$			&	882$\pm$	17		&	1.05/37(0.38)\\
91445009$^a$	&	56129.1	&	1.8$\pm$?	&	10440$\pm?$			&	4333$\pm$?		&     3.94/56($1.9$$\times$$10^{-21}$)\\
91445010		&	56134.1	&	1.62$\pm$0.05	&	16070$\pm330$		&	6669$\pm$137		&	0.77/36(0.84)\\
13708$^a$(CXO)&	56138.4	&	1.68$\pm$0.03&	6680$\pm80$			&	2772$\pm$33		&     1.76/102($3.4$$\times$$10^{-6}$)\\
91445011$^a$	&	56140.1	&	1.8$\pm$?	&	7330$\pm?$			&	3042$\pm$?		&     2.10/80($3.4$$\times$$10^{-6}$) \\
91445012		&	56144.8	&	2.11$\pm$0.03	&	3651$\pm70$			&	1515$\pm$29		&	1.12/38(0.27) \\
91445013		&	56149.4	&	2.07$\pm$0.07	&	514$\pm20$			&	213$\pm$	8		&	1.10/58(0.28) \\
32148008		&	56150.1	&	1.7$\pm$0.1	&	295$\pm16$			&	122$\pm$	7		&     0.81/45(0.81) \\
32148011		&	56152.2	&	1.8$\pm$0.1	&	260$\pm16$			&	108$\pm$7		&     0.70/37(0.91)\\
530808000	&	56152.4	&	1.9$\pm$0.2	&	266$\pm30$			&	110$\pm$	12		&	1.12/10(0.34) \\
32148010$^a$	&	56153.2	&	1.8$\pm$0.1	&	219$\pm13$			&	91$\pm$5			&	1.79/44($9.5$$\times$$10^{-4}$)\\
91445014		&	56154.3	&	1.9$\pm$0.2	&	215$\pm18$			&	89$\pm$7			&	1.06/23(0.38) \\
32148012		&	56158.3	&	2.0$\pm$0.1	&	149$\pm11$			&	62$\pm$5			&	1.06/33(0.37) \\
91445015		&	56159.1	&	1.8$\pm$0.3	&	110$\pm13$			&	46$\pm$5			&	0.97/10(0.47) \\
91445016		&	56163.2	&	2.0$\pm$0.2	&	45$\pm5$				&	19$\pm$2			&	1.49/16(0.09)\\
32148013		&	56169.3	&2.4$_{-1.2}^{+1.5}$&	2.7$_{-1.5}^{+4.8}$		&	1.1$_{-0.6}^{+1.9}$	&	-\\
\hline
\end{tabular}
\end{center}
\caption{\Swiftxrt\ and \Chandraacis\ observations fit to an absorbed power-law model. Unabsorbed flux in units of $10^{-12}$erg s$^{-1}$ cm$^{-2}$ and Luminosity in $10^{34}$ erg s$^{-1}$, both in 0.5 - 10 keV. all the flux and luminosity values are background-subtracted. ``a" indicates spectra which are not well-described by a power-law (null hypothesis probability  is $< 10^{-2}$); if $\chi^2 > 2$, no errors are calculated (indicated by ``?"). The first row shows a spectral fit to a merged spectrum of 6 \Swiftxrt\ observations of Terzan 5 before the outburst of Terzan 5 X-3. For Obs.ID 32148013, due to low counts, we used \emph{cstat} statistics in the fitting. Reported uncertainties are 90$\%$ intervals.}
\label{powerlaw_data}
\label{pre-outburst-table}
\end{table*}

\subsubsection{Thermonuclear burst}\label{thermo}
\begin{figure*}[h]
\begin{center}
\includegraphics[scale=0.42]{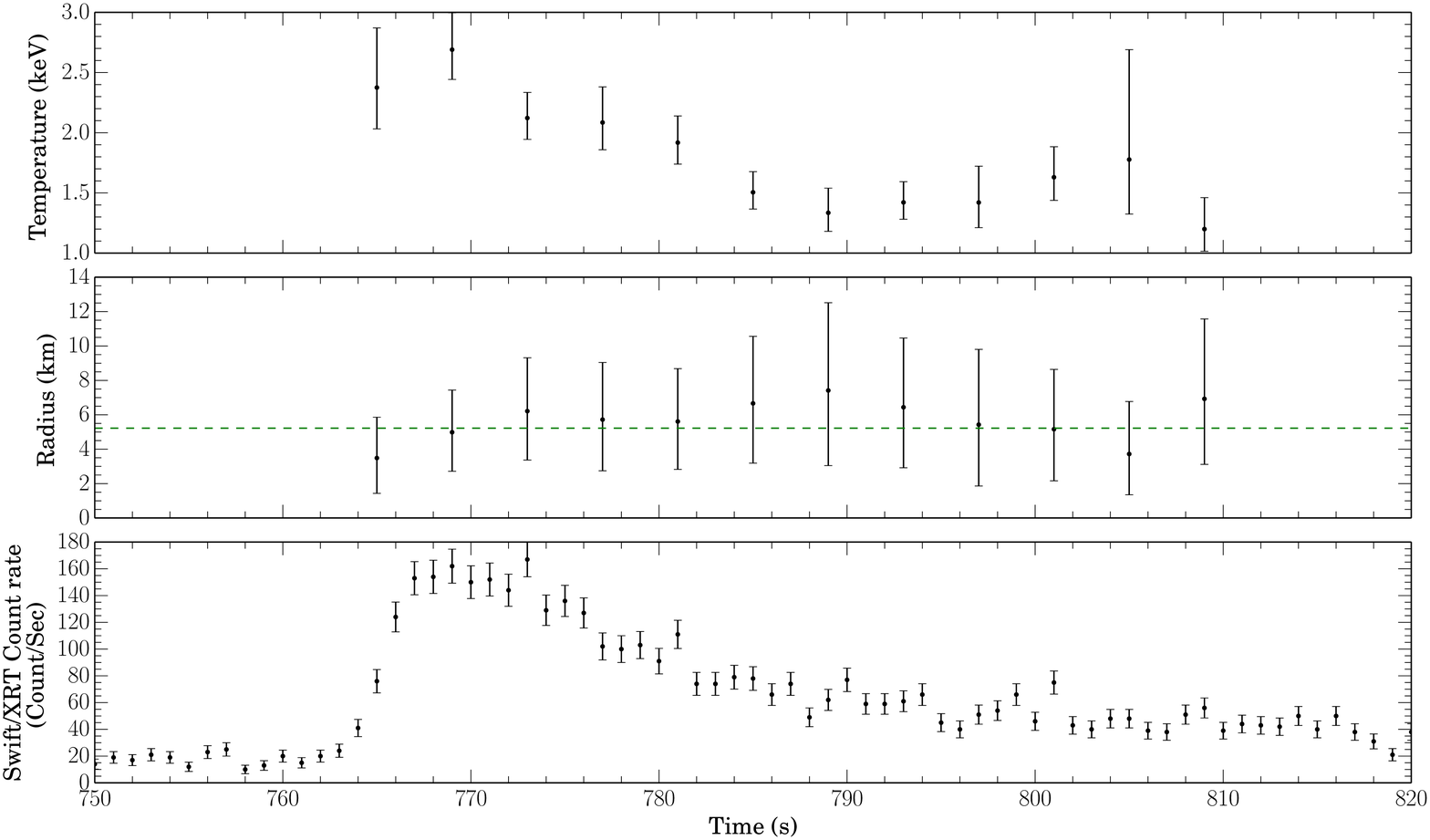}
\caption{Spectral evolution during the thermonuclear burst of Terzan 5 X-3, using blackbody fits. The first and second panels show the variation in temperature and radius respectively. The third panel is the 0.5-10 keV light curve of the burst. The significant drop of temperature around 770 - 790 s shows evidence for cooling during the thermonuclear burst. The green line in the second panel represents the weighted average radius. Error bars are 1-$\sigma$ uncertainties. Time bins in the top two panels are 4s long.}
\label{thermoplot}
\end{center}
\end{figure*}

During \Swiftxrt\ ObsID 32148007{, which started at 20:54:00 UT on 2012 July 17} we detected an eightfold count rate increase over $\sim$ three seconds, { starting at $\approx$ 21:06:40}, followed by a slower decline over $\sim$1 minute (Fig.~\ref{thermoplot}, bottom panel), suggestive of a thermonuclear burst \citep{Altamirano12atel}. Using FTOOLS \emph{xselect}, we divided the data from this observation into time intervals of {4 seconds each}. We extracted spectra from each time interval using FTOOLS \emph{xselect} and analyzed the spectra. 

During this thermonuclear burst, the count rate reached $\approx$ 160 count s$^{-1}$. There is a possibility of pileup in \Swiftxrt\ observations in windowed timing mode when the count rate exceeds 100 count s$^{-1}$ (\citealt{Romano06}). {Following the \Swiftxrt\ pile-up thread for WT data, we extracted spectra{ while excluding} increasingly large fractions of the central PSF. Doing this, we found that the fitted photon index did not change, and thus we found no significant signs of pileup in this observation during the burst.}

We fit an absorbed power-law to the spectrum extracted from a pre-burst interval, and considered this fit as a fixed component of the spectral model for time intervals during the burst (cf., \citealt{Worpel13}; our statistics are insufficient to determine whether this assumption is correct, and moderate changes should not affect our conclusions).  We fit an absorbed blackbody model (BBODYRAD) to the burst emission, finding decent fits for all intervals, and show the spectral evolution in Fig.~\ref{thermoplot}. {We find clear evidence of cooling (between 770 - 790 s in Fig.~\ref{thermoplot})}, while the inferred radius remains essentially constant. This is a clear signature of a thermonuclear burst, and thus of a NS.

We estimated the timescale of this burst using two methods. We fit an exponential model (count rate $\propto e^{-t/\tau_1}$) to the light curve of the burst after the peak, estimating the timescale $\tau_1\approx 16\pm1 $s.  Following \cite{Galloway08}, we estimated an alternative timescale for thermonuclear bursts {$\tau_2 = E_{burst} / F_{peak} $},{ where $E_{burst}$ is the total fluence during the burst and $F_{peak}$ is flux at the peak, finding} $\tau_2 \approx 29 $s. \citet{Galloway08} divide bursts into those with $\tau_2$ longer than 10 s, and those with $\tau_2$ shorter than 10 s.  The longer bursts are generally powered by hydrogen burning (with the exception of ``giant'' bursts involving photospheric radius expansion, which was not seen here), and the short bursts involve only helium burning, since 
hydrogen burning proceeds more slowly than helium burning \citep{Fujimoto81,vanParadijs88,Cornelisse03}. Our measured burst timescale indicates that hydrogen is being accreted, and thus that the donor star is hydrogen-rich, which requires an orbital period $\gtrsim$1.5 hr (e.g., \citealt{Nelson86}) and excludes a WD donor.

\newpage
\subsubsection{Quiescent behaviour}\label{sec:quiescence}
\begin{figure*}[h]
\begin{center}
\includegraphics[scale=0.42]{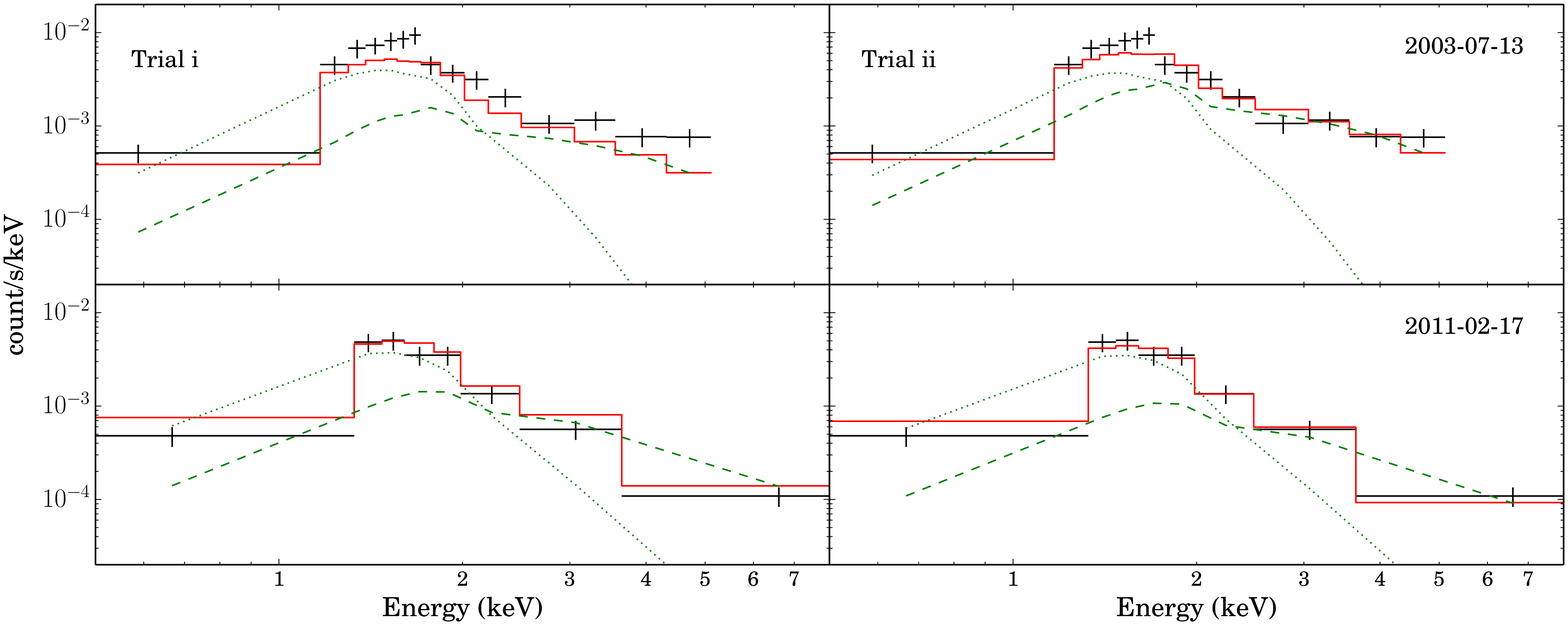}
\caption{Extracted spectra from two of the \Chandraacis\ observations of Terzan 5 X-3, compared to the fitted model (NSATMOS + PEGPWRLW in XSPEC) from Trials I and II (Table \ref{qui_fit_details}).  The dotted line indicates the contribution of the NSATMOS, while the dashed line indicates the contribution from the power-law component, and the solid line is their sum. Spectral variations are clear. Comparing Trial II to I fitting improvement can be seen.}
\label{chandra_spec}
\end{center}
\end{figure*}

We used 7 \Chandraacis\ observations taken when all sources were quiescent (Table \ref{chandra_data}) to study the behaviour of Terzan 5 X-3 in quiescence before its outburst. We extracted source and background spectra from each observation using CIAO \emph{dmextract}. {We used a combination of a power-law (PEGPWRLW) model and a hydrogen atmosphere for a NS (NSATMOS), with absorption (PHABS)} set to our preferred cluster value, the NS radius to 10 km, mass to 1.4 \Msun, and distance set to 5.9 kpc. This model has been previously used to fit its spectrum in one quiescent observation \citep{Heinke06b}. 
 To study {possible }spectral variations, we simultaneously fit spectra from each observation in four different Trials, each with different parameters free (Table \ref{qui_fit}) : 
I) constraining the NSATMOS and PEGPWRLW components to have the same values between all observations; 
II) letting only the power-law normalization vary between observations, while constraining the  NSATMOS temperature to be the same; 
III) letting only the NSATMOS temperature vary between observations, while constraining the power-law normalization to be the same; 
IV) letting both the power-law normalization and the NSATMOS temperature vary between observations.
We found no evidence for variation in the power-law photon index $\Gamma$ between observations if we allowed it to vary. Therefore, we tied its value between observations in each Trial (Fig.~\ref{chandra_spec}).  

{Trial I gives a poor fit, with $\chi^2$=116.3 for 76 D.O.F. An F-test confirms the improvement from allowing the power-law flux to vary (Trial I to II), giving an F-statistic of 6.0 and probability of $4\times10^{-5}$ of obtaining such an improvement by chance.  Alternatively, allowing the NS temperature to vary (Trial III) also gives an improvement compared to Trial I (F-statistic of 4.2, chance improvement probability of $1\times10^{-3}$).  Letting both components vary is a substantial improvement compared to allowing the NS temperature alone to vary (comparing III to IV, F-statistic=2.27, chance improvement probability of $4.7\times10^{-2}$), while letting both components vary is not preferred over allowing the power-law component alone to vary (comparing II to IV, F-statistic=0.89, chance improvement probability 0.50).  Thus, we identify clear variation in the non-thermal component, but no evidence for variation in the thermal component.}

\begin{table*}[h]
\begin{center}
\begin{tabular}{lccccccc}
\hline
\hline
	&		&	log T							&	$F_{X,NS}$(0.5-10 keV)			&					&	$F_{X,PL}$(0.5-10 keV)		&$L_{X,total}$(0.5-10 keV) &	\\
Trial & ObsID	&	(K)							&(10$^{-13}$ erg cm$^{-2}$ s$^{-1}$)	&	$\Gamma$		&(10$^{-13}$ erg cm$^{-2}$ s$^{-1}$)&(10$^{33}$ erg s$^{-1}$)&	$\chi_\nu^2$/D.O.F(nhp)\\
\hline
I & all		&	6.143$_{-0.011}^{+0.015}$		&		2.4$_{-0.2}^{+0.3}$			&	1.9$\pm0.4$		&	1.1$_{-0.1}^{+0.3}$			&	1.4$_{-0.1}^{+0.2}$	&	1.53/76(0.002)		\\
\hline	
II & 03798		&	6.138$_{-0.018}^{+0.012}$		&		2.4$_{-0.5}^{+0.3}$			&	2.1$\pm0.4$		&	1.9$_{-0.4}^{+0.6}$			&	1.8$\pm0.3$		&	1.09/70(0.27)		\\
 & 10059		&	t							&		t						&	t				&	1.3$_{-0.3}^{+0.5}$			&	1.5$\pm0.2$		&	t				\\ 
 & 13225		&	t							&		t						&	t				&	0.8$_{-0.3}^{+0.4}$			&	1.3$\pm0.2$		&	t				\\
 & 13252		&	t							&		t						&	t				&	1.2$_{-0.3}^{+0.5}$			&	1.5$\pm0.2$		&	t				\\ 
 & 13705		&	t							&		t						&	t				&	0.7$\pm0.6$				&	1.3$\pm0.3$		&	t				\\
 & 14339		&	t							&		t						&	t				&	0.8$_{-0.3}^{+0.5}$			&	1.3$\pm0.2$		&	t				\\
 & 13706		&	t							&		t						&	t				&	1.4$_{-0.3}^{+0.4}$			&	1.6$\pm0.2$		&	t				\\
\hline
III & 03798	&	6.172$_{-0.015}^{+0.011}$		&		3.3$_{-0.5}^{+0.4}$			&	1.9$\pm0.4$		&	1.0$_{-0.2}^{+0.3}$			&	1.8$\pm0.2$		&	1.22/70(0.099)		\\
 & 10059		&	6.149$_{-0.019}^{+0.014}$		&		2.7$_{-0.5}^{+0.3}$			&	t				&	t						&	1.5$\pm0.2$		&	t				\\ 
 & 13225		&	6.120$_{-0.029}^{+0.019}$		&		1.9$\pm0.5$				&	t				&	t						&	1.2$\pm0.2$		&	t				\\
 & 13252		&	6.144$_{-0.019}^{+0.014}$		&		2.4$_{-0.5}^{+0.3}$			&	t				&	t						&	1.4$\pm0.2$		&	t				\\ 
 & 13705		&	6.115$_{-0.042}^{+0.025}$		&		1.9$_{-0.7}^{+0.5}$			&	t				&	t						&	1.2$\pm0.3$		&	t				\\
 & 14339		&	6.131$_{-0.023}^{+0.017}$		&		2.2$_{-0.5}^{+0.2}$			&	t				&	t						&	1.3$\pm0.2$		&	t				\\
 & 13706		&	6.155$_{-0.017}^{+0.012}$		&		2.7$_{-0.5}^{+0.3}$			&	t				&	t						&	1.5$\pm0.2$		&	t				\\
\hline
IV & 03798       	&	6.161$_{-0.021}^{+0.015}$		&		3.0$_{-0.6}^{+0.3}$			&	1.7$\pm0.4$		&	1.5$\pm0.4$				&	1.9$_{-0.3}^{+0.2}$	&	1.10/64(0.27)		\\
 & 10059		&	6.148$_{-0.021}^{+0.016}$		&		2.7$_{-0.5}^{+0.6}$			&	t				&	1.1$_{-0.3}^{+0.4}$			&	1.6$_{-0.2}^{+0.3}$	&	t				\\
 & 13225		&	6.134$_{-0.024}^{+0.015}$		&		2.2$_{-0.5}^{+0.2}$			&	t				&	0.7$_{-0.2}^{+0.3}$			&	1.2$\pm0.2$		&	t				\\
 & 13252		&	6.146$_{-0.021}^{+0.015}$		&		2.7$_{-0.5}^{+0.3}$			&	t				&	1.0$_{-0.3}^{+0.4}$			&	1.5$\pm0.2$		&	t				\\
 & 13705		&	6.121$_{-0.043}^{+0.025}$		&		1.7$\pm1.0$				&	t				&	1.3$\pm0.9$				&	1.2$\pm0.5$		&	t				\\
 & 14339		&	6.142$_{-0.019}^{+0.014}$		&		2.4$_{-0.5}^{+0.3}$			&	t				&	0.7$_{-0.3}^{+0.4}$			&	1.3$\pm0.2$		&	t				\\
 & 13706		&	6.154$_{-0.021}^{+0.014}$		&		2.7$_{-0.5}^{+0.3}$			&	t				&	1.1$_{-0.2}^{+0.3}$			&	1.6$\pm0.2$		&	t				\\
\hline
\end{tabular}
\end{center}
\caption{Spectral fitting of 7 \Chandraacis\ observations of Terzan 5 X-3 during quiescence (see Table \ref{chandra_data}), using PHABS(PEGPWRLW+NSATMOS) in XSPEC. In Trial I both components are constrained to have the same values between observations. In Trials II and III one of the components may vary between observations, while in the 4th Trial both components are free. We use a ``t" whenever values of a parameter are tied between observations.
 kT is the NS surface temperature{ in the star's frame}, $\Gamma$ is the power-law photon index, $F_{X,NS}$ is the unabsorbed flux from the NS atmosphere component, and $F_{X,PL}$ is the unabsorbed flux from the power-law component. {Uncertainties are 90$\%$ confidence intervals. nhp is the null hypothesis probability (otherwise known as the p-value).}}
\label{qui_fit_details}
\label{qui_fit}
\end{table*}

\subsubsection{Rise of the outburst}\label{sec:rise}
We fit the \Swiftxrt\ spectra  from the rise of the outburst with a two-component model including a thermal component (BBODYRAD in XSPEC) and a non-thermal component (PEGPWRLW in XSPEC).   We found good fits permitting only the relative normalizations of the thermal and non-thermal components to vary, with a photon index tied between observations ($\Gamma$=1.1$^{+0.2}_{-0.4}$) and a blackbody radius tied between observations (R=4.3$^{+1.4}_{-1.2}$ km). When the power-law index is left free between observations, the values are consistent, though they are poorly constrained in several spectra.
Comparing our two-component model fits (Table \ref{rise_thermo}) to our power-law fits (Table \ref{powerlaw_data}), a clear improvement in the fit is seen.
Simultaneous fits to the first five \Swiftxrt\ spectra of the outburst (listed in Table \ref{rise_thermo}) with an absorbed power-law (with the photon index free between observations) give a reduced $\chi^2$ of 1.32 for 45 degrees of freedom, while fits with an absorbed power-law plus blackbody give a reduced $\chi^2$ of 1.09 for 43 degrees of freedom.  An F-test gives an F-statistic of 5.55 and chance improvement probability of 0.007, supporting the addition of the thermal component. \citet{Protassov02} showed that the F-test is often inaccurate for testing the necessity of adding an additional spectral component.  We therefore chose the spectrum with the clearest evidence of a thermal component (ObsID 32148004), which shows a $\Delta \chi^2$ of 6.1 between the power-law and power-law plus thermal spectral fits (going from 12 degrees of freedom to 11).  We simulated 1000 data sets using a best-fit absorbed power-law model, and fit them both with a power-law model and with a power-law plus thermal component model. None of our simulations showed a larger $\Delta \chi^2$ than that produced by our model, allowing us to conclude that the probability of incorrectly concluding that a thermal component is required is less than 99.5\%.

This indicates that the hardening during the outburst rise is likely caused by the decreasing relative contribution of a thermal component. With increasing time, and thus with increasing $L_x$, the fractional contribution of the thermal component decreases, but its kT increases monotonically. In the next section, we suggest that the thermal component is due to low-level accretion onto the surface of the neutron star.

\section{Discussion}\label{sec:disc}
\subsection{Hardening during the outburst rise}
We observed clear evidence of hardening of the spectrum during the outburst rise from $L_X\sim4\times10^{34}$ up to $10^{36}$ ergs/s.  We have evidence that this hardening is due to the relative reduction in strength of a thermal component in the spectrum with increasing brightness.  
This is the first time that such hardening during the outburst rise has been detected, made possible by our program of \Swiftxrt\ globular cluster monitoring allowing early detection of the outburst below $L_X=10^{35}$ ergs/s.  The trend of inferred photon index (for a fit to a power-law model) versus $L_X$ is clear from the data in the rise, and is consistent with the data in the decay (which are not well-sampled below $L_X=10^{35}$ ergs/s); see Fig.~\ref{spec_ev}.

Softening during outburst decays has been seen from other (likely) NS LMXBs, in the  $L_X$ range of $10^{34}-10^{35.5}$ ergs/s, especially when the soft ($<$2 keV) X-ray energy range is included \citep{Jonker03,Jonker04b,Cackett11,Fridriksson11,ArmasPadilla11,ArmasPadilla13c}.  \RXTE\ observations have  shown marginal softening during the decay of Aql X-1 down to $L_X=5\times10^{34}$ ergs/s \citep{Maitra04}, only in the part of the spectrum below 6 keV.  \RXTE\ observations of SAX J1808.4-3658 showed almost no spectral changes from $L_X\sim2\times10^{36}$ down to $2\times10^{34}$ ergs/s \citep{Gierlinski98}.  These apparently contrasting observations are consistent if the softening is due to the increasing importance of the thermal component at lower $L_X$.  Evidence in favor of an increasing relative thermal component can also be seen in \Swiftxrt\ spectra of SAX J1808.4-3658 declining from $L_X\sim10^{36}$ down to $10^{33}$ ergs/s \citep{Campana08}.  Thus, we interpret the hardening we observe in Terzan 5 X-3's rise as due primarily to the decreasing importance of a thermal component, rather than to the same physics responsible for the softening of black hole LMXBs during their decay, which show a steepening power-law spectrum \citep[e.g.,][]{Plotkin13}.

Comparing the spectra observed from Terzan 5 X-3 to those of other NS LMXBs, we find a common pattern, that below $L_X\sim1-3\times10^{35}$ ergs/s a thermal component is often required.  For instance, \citet{ArmasPadilla13a} find, using XMM-Newton spectra, that two LMXBs at $L_X=1-10\times10^{34}$ ergs/s require a strong thermal component, while this is not critical for another LMXB at $L_X\sim10^{35}$ ergs/s.  \citet{Wijnands02d}, using \Chandra\, find that SAX J1747.0-2853, at $L_X\sim3\times10^{35}$ ergs/s, does not need a thermal component.  \citet{ArmasPadilla13c} measure a thermal component to comprise $\sim$20\% of the 0.5-10 keV luminosity for a transient at $L_X\sim9\times10^{34}$ ergs/s (using XMM), with no evidence (from poorer \Swiftxrt\ spectra) for a thermal component above $L_X=2.6\times10^{35}$ ergs/s. \citet{Jonker03,Jonker04b} study the return to quiescence of XTE J1709-267, finding a thermal component to comprise $\sim40$\% of the flux at $L_X\sim4\times10^{34}$ ergs/s, increasing to $>90$\% at $2\times10^{33}$ ergs/s.  
All these results suggest that there is a physical transition operating around $L_X\sim10^{35}$ ergs/s which changes the energy spectra. 

\begin{table*}[h]
\begin{center}
\begin{tabular}{cccccccc}
\hline
\hline
			&			&					&	F$_{X,BB}$ (0.5-10 keV)			&		F$_{X,PL}$ (0.5-10 keV )		&						&	L$_{X,total}$(0.5-10 keV)	&		\\
Obs. ID 		&       MJD		& 	kT (keV) 			&(10$^{-12}$ erg s$^{-1}$ cm$^{-2}$)	&(10$^{-12}$ erg s$^{-1}$ cm$^{-2}$)	&F$_{X,PL}$/F$_{X,total}$	&	(10$^{34}$ erg s$^{-1}$)	& $\chi_\nu^2$/D.O.F\\
\hline
91445006		&	56114.8	&	0.31$\pm$0.03		&	5$\pm$2						&		5$\pm$2					&	50$\pm 20 \%$			&	4$\pm1$				&	0.53/6\\
32148003		&	56115.8	&	0.36$\pm$0.03		&	9$\pm$3						&		13$\pm$4					&	59$_{-16}^{+15} \%$	&	9$\pm2$				&	0.68/5\\
32148004		&	56117	&	0.41$\pm$0.02		&	15$_{-3}^{+4}$					&		17$\pm$6					&	53$_{-16}^{+12} \%$	&	13$\pm3$				&	1.19/9\\
32148005		&	56118.1	&0.44$_{-0.07}^{+0.05}$	&	20$\pm$10					&		70$\pm$20				&	78$_{-15}^{+12} \%$	&	37$\pm9$				&	0.55/6\\
32148006		&	56120.7	&	0.67$\pm$0.06		&	110$\pm$40					&		500$_{-70}^{+60}$			&	82$_{-8}^{+7} \%$		&	250$\pm30$			&	1.39/19\\
\hline
\end{tabular}
\end{center}
\caption{Results of spectral analyses for the rise of the outburst with a two-component model: thermal (BBODYRAD in XSPEC) plus non-thermal (PEGPWRLW in XSPEC). F$_{X,BB}$ and F$_{X,PL}$ are the blackbody and power-law unabsorbed fluxes, respectively. The power-law photon index was tied between spectra, and found to be 1.1$_{-0.4}^{+0.2}$. The thermal component normalization (which is proportional to blackbody radius) is assumed constant, and tied between observations. Uncertainties are 90$\%$ confidence intervals. $\chi^2_\nu$ and degrees of freedom in this table are found by fitting each dataset individually based on values found in simultaneous fit.}
\label{rise_thermo}
\end{table*}

Such a transition can be provided by the declining optical depth of a hot Comptonizing atmosphere, as seen in numerical calculations of NSs accreting at low rates \citep{Deufel01,Popham01}.  Deufel et al.\ 2001 show temperature profiles and emergent spectra for NSs illuminated by high-temperature protons, such as are produced by radiatively inefficient accretion flows. 
in their figs. 4 and 5 they show that the emergent spectrum is a featureless Comptonized spectrum extending to $\sim$100 keV above $L_X\sim10^{36}$ ergs/s, which develops a clear 0.5 keV thermal component at $\sim10^{35}$ ergs/s, and loses the Comptonized tail by $10^{33}$ ergs/s. This transition is a strikingly accurate match to our observations of Terzan 5 X-3's spectral variations, and to other NS transients discussed in the literature. However calculations of Deufel et al.\ 2001 underpredict the observed hard power-law components seen in many quiescent LMXBs at low $L_X$ ($<10^{34}$ ergs/s, including Terzan 5 X-3 in quiescence). This may arise from their not including the Comptonizing effects of the overlying accretion flow on the observed spectrum.  \citet{Popham01} compute solutions for a hot boundary layer, which becomes optically thin for accretion luminosities below $\sim10^{36}$ ergs/s, suggesting that some additional Comptonization can be performed by the accretion flow.  

The temperature of the thermal component increases monotonically with the total X-ray luminosity during the rise (Table 4), as expected if the thermal component during the rise is produced by accretion.  A correlation of thermal component temperature with total luminosity has been suggested from comparisons of multiple sources \citep{ArmasPadilla13c}, but this measurement clearly confirms this correlation in a single source.  
Furthermore, heating of the NS crust by accretion during the outburst will give rise to a rapidly decaying surface temperature at the end of the outburst \citep[e.g.,][]{Cackett06b}. This effect of an accretion-heated crust could be confused with changing thermal emission from low-level accretion onto the NS surface during the outburst decline, but is not an issue during the outburst rise.

\subsection{Nature of donors in globular cluster X-ray binaries}
The detection of a thermonuclear burst during this outburst showed that the accreting object is a NS. Furthermore, the timescale of this thermonuclear burst indicates that the accreted matter contains hydrogen, evidence that the donor is not a white dwarf.  With this information, we are now capable of classifying 15 of the 18 known bright Galactic globular cluster LMXBs as either ultracompact ($P_{orb}<$ 1 hour, accreting from a hydrogen-deficient and/or degenerate donor) or not ultracompact (accreting from a nondegenerate, H-rich star).  
Five are known to be ultracompact by direct detection of their orbital periods, and seven systems are known not to be ultracompact by direct measurement of their orbital periods. On the basis of X-ray burst behaviour indicative of hydrogen burning \citep{Galloway08}, another three systems can be identified as not ultracompact (Table \ref{lmxbs}).  (4U 1722-30, in Terzan 2, has shown some evidence, by its persistent low-luminosity accretion and burst behaviour, in favor of an ultracompact nature; \citealt{intZand07}.)  Thus, the fraction of observed bright globular cluster LMXBs that are ultracompact appears to be between 28-44\% (for 5 or 8 of 18). This fraction is believed to be higher than in the rest of the Galaxy \citep{Deutsch00}, but uncertainties in selection effects mean that we cannot confidently extrapolate the true underlying fraction of ultracompact systems and make clear comparisons to binary population synthesis models \citep[e.g.,][]{Ivanova08}.

\subsection{Quiescent counterpart}
We have identified the quiescent counterpart to Terzan 5 -X3 with the brightest previously suggested candidate quiescent LMXB in the cluster, CXOGLB J174805.4-244637 \citep{Heinke06b}.  
Our spectral analysis reveals evidence for a variable power-law contribution to the quiescent spectrum over timescales of years, but exhibits no evidence for changes to the thermal component.  It is fascinating to see clearly here that the quiescent spectral properties appear to lie on a continuum with the outburst properties, with increasing hardening from quiescence, through the early rise, up to the hard state at $L_X>10^{36}$ ergs/s (Fig.~\ref{spec_ev}). 

\Chandra\ quiescent X-ray counterpart searches have now been performed for nine transient cluster LMXBs, 
of which the three with the faintest outbursts have been identified with very faint ($L_X<10^{32}$ ergs/s) quiescent counterparts 
(M15 X-3, \citealt{Heinke09b}; NGC 6440 X-2, \citealt{Heinke10}; IGR J17361-4441 in NGC 6388, \citealt{Pooley11}), 
two have spectrally hard counterparts with $L_X\sim10^{32-33}$ ergs/s 
(EXO 1745-248 in Terzan 5, \citealt{Wijnands05}; IGR J18245-2452 in M28, \citealt{Papitto13,Linares13}),  and four have spectrally soft counterparts with $L_X\sim10^{32-33}$ ergs/s 
(SAX J1748.9-2021 in NGC 6440, \citealt{intZand01}; X1732-304 in Terzan 1, \citealt{Cackett06a};  IGR J17480-2446 in Terzan 5, \citealt{Degenaar11b}; and Terzan 5 X-3).
These identifications support the idea that the faint soft X-ray sources identified as candidate quiescent LMXBs in globular clusters are indeed transient LMXBs, between (relatively bright) outbursts \citep{Heinke03d,Wijnands13}.  
The brightest of the faint soft X-ray sources should experience relatively high long-term average mass accretion rates, which will cause relatively large amounts of deep crustal heating and keep their cores warm.  
Such high mass accretion rates suggest frequent outbursts, and thus it is comforting that we have identified the brightest of the faint soft X-ray sources in the clusters NGC 6440, Terzan 5, and Terzan 1 with observed transients.
 The suggestion that roughly half the quiescent LMXBs in each cluster are easily identifiable in short \Chandra\ observations by showing soft, primarily thermal X-ray spectra and X-ray luminosities between $10^{32-33}$ ergs/s \citep{Heinke05b} continues to seem reasonable, though it remains unproven.

From the quiescent NS thermal bolometric luminosity ($L_{NS, (0.01-10 keV)}=1.5\times10^{33}$ ergs/s, kT=118 eV at the surface) of Terzan 5 X-3, and its outburst properties, we can make some general statements about its outburst history or neutrino cooling properties, assuming that the quiescent thermal flux is due to heat deposited in the core during multiple accretion episodes \citep{Brown98}.  We estimate the total mass transfer rate onto the NS during this outburst by converting the daily \Maxigsc\ 4-10 keV flux estimates (in Crab units) into 0.1-12 keV fluxes (assuming a power-law with photon index set by the nearest \Swiftxrt\ observations), converting the daily \Swiftbat\ flux estimates into 12-50 keV fluxes (using the same power-law photon index as for the \MAXI\ data), adding these together, and assuming a 1.4 \Msun, 10 km NS.  This gives us a total energy release over the outburst of 9$\times10^{43}$ ergs, and a total mass transfer of 2.4$\times10^{-10}$ \Msun. 

If we assume ``standard'' modified Urca cooling {\citep{Yakovlev04,Wijnands13}}, then we can estimate (using the quiescent NS luminosity) a mass transfer rate onto the NS of \.{M}$\sim3\times10^{-11}$ \Msun/year (though this value might vary depending on the choice of crustal composition, e.g. whether a thick light-element layer is present; \citealt{Page04}).  Assuming this outburst is typical, we derive an average interval of $\sim$8 years between outbursts.  If the average interval between outbursts were much {longer} than 10 years, then Terzan 5 X-3 would be brighter in quiescence than expected under even the slowest cooling processes.  One could attribute its quiescent thermal luminosity to continued accretion, but our analysis of the quiescent observations identifies no evidence for variability in the thermal component, arguing against this explanation. 
 All Terzan 5 X-ray outbursts since 1996 have been identified with their quiescent counterpart with arcsecond precision, except one in 2002 \citep{Wijnands02Atel}.
 The 2002 outburst had an average luminosity of $L_X\sim 2\times10^{37}$ erg s$^{-1}$, peak $L_X\sim4\times10^{37}$ erg s$^{-1}$, and lasted for $\sim 30$ days \citep{Degenaar12}. The 2012 outburst of Terzan 5 X-3 had a similar average luminosity of $L_X\sim 2\times10^{37}$ erg s$^{-1}$, peak $L_X=7\times10^{37}$ erg s$^{-1}$, and lasted for 30 days above $L_X\sim10^{36}$ ergs s$^{-1}$ (comparable to the RXTE/ASM detection limit for the 2002 outburst). We therefore suggest that the 2002 X-ray outburst is likely to have also been produced by Terzan 5 X-3. This would nicely fit the $\sim$8 year recurrence time we inferred above. 

\begin{table*}[h]
\begin{center}
\begin{tabular}{lcccccc}
\hline
\hline
LMXB	 		& globular cluster & State & P$_{orb}$ & nature & notes & references\\
\hline
4U 1820-30 		& NGC 6624	& P	& 11 min 		& U & X & (1)\\
4U 0513-40 		& NGC 1851	& P	& 17 min		& U & UV & (2)\\
X1850-087  		& NGC 6712	& P	& 20.6$^a$ min & U & UV & (3)\\
M15 X-2 	    		& M 15 	 	& P	& 22.6 min	& U & UV & (4)\\
NGC 6440 X-2 		& NGC 6440	& T	& 57.3 min	& U & XP & (5)\\ 
XB 1832-330 		& NGC 6652 	& P	& 2.1 hrs 		& N & O & (6)\\ 
4U 1746-37 		& NGC 6441 	& P	& 5.16 hrs 	& N & X & (7)\\
SAX J1748.9-2021 	& NGC 6440 	& T	& 8.7 hrs 		& N & XP & (8)\\ 
IGR J18245-2452 	& M28 	 	& T	& 11.0 hrs		& N & XP & (9)\\
GRS 1747-312 	& Terzan 6 	& T	& 12.36 hrs 	& N & X & (10)\\
AC 211 			& M 15  		& P	& 17.1 hrs 	& N & UV & (11)\\
Terzan 5 X-2 		& Terzan 5 	& T	& 21.27 hrs	& N & XP & (12,13)\\
Rapid Burster		& Liller 1		& T	& ?			& N & B & (14) \\	
EXO 1745-248	& Terzan 5 	& T	& ?			& N & B & (14) \\
Terzan 5 X-3		& Terzan 5 	& T	& ?			& N & B & (15)\\
XB 1732-304& Terzan 1&T& ?			& ?  & ? & (16)\\
4U 1722-30& Terzan 2 &P & ?			&U?& B & (17)\\
IGR J17361-4441& NGC 6388 & T& ?  	& ?  & ? & (18)\\
\hline
\multicolumn{2}{l}{LMXBs with no observed outbursts} \\
\hline
47 Tuc W37		& 47 Tuc		& Q	& 3.09 hrs		& N & X & (19) \\
47 Tuc X5			& 47 Tuc		& Q	&  8.67 hrs	& N & X & (20) \\
$\omega$ Cen qLMXB & $\omega$ Cen & Q & ? 		& N & H$\alpha$ & (21) \\
\hline
\end{tabular}
\end{center}
\caption{Orbital periods, or other classification, of the 18 Galactic globular cluster LMXBs that are persistently bright or have shown luminous outbursts, plus three quiescent globular cluster LMXBs. X-ray bursts have been detected from all the bright LMXBs except for AC 211 in M15 and IGR J17361-4441 in NGC 6388.  State: P=persistent (or active for $>$30 years), T=transient, Q=quiescent (so far).
Nature: U=ultracompact, N=normal.  Notes: X=period from X-ray photometry, UV=period from UV photometry, XP=period from X-ray pulsations, O=period from optical photometry, B=nature of donor inferred from properties of X-ray bursts, H$\alpha$=hydrogen seen in optical counterpart spectrum. Notes represent method of measuring P$_{orb}$ or determining donor natures. References: 1- \citealt{Stella87}, 2- \citealt{Zurek09}, 3- \citealt{Homer96},  4- \citealt{Dieball05}, 5- \citealt{Altamirano10}, 6- \citealt{Engel12}, 7- \citealt{BalucinskaChurch04}, 8- \citealt{Altamirano08}, 9- \citealt{Papitto13}, 10- \citealt{intZand03}, 11- \citealt{Ilovaisky93},  12- \citealt{Strohmayer10Atel2}, 13- \citealt{Papitto11},  14- \citealt{Galloway08}, 15- this work, {16-\citealt{Guainazzi99}, 17- \citealt{intZand07}, 18- \citealt{Bozzo11}, }19- \citealt{Heinke05b}, 20- \citealt{Heinke03a}, 21- \citealt{Haggard04}. a- Or the alias period of 13 minutes.  }
\label{lmxbs}
\end{table*}

\acknowledgements
We thank H. A. Krimm for helpful discussion on the analysis of \Swiftbat\ survey data, and T. Mihara for helpful discussion on \Maxigsc\ calibration issues. 
 We acknowledge financial support from NSERC Discovery Grants (C.O.H., G.R.S.), an Alberta Ingenuity New Faculty Award (C.O.H.) and the Avadh Bhatia Fellowship (J.C.G.). {ND is supported by NASA through Hubble Postdoctoral Fellowship grant number HST-HF-51287.01-A from the Space Telescope Science Institute. RW is supported by an European Research Council Starting Grant. DA acknowledges support from the Royal Society. J.H. and D.P. acknowledge support by the National Aeronautics and Space Administration through Chandra Award Number GO2-13045B, issued by the Chandra X-ray Observatory Center, which is operated by the Smithsonian Astrophysical Observatory for and on behalf of the National Aeronautics Space Administration under contract NAS8-03060.}

This research has made use of the following data and software packages: observations made by the Chandra X-ray Observatory, data obtained from the Chandra Data Archive, software provided by the Chandra X-ray Center (CXC) in the application package CIAO, \MAXI\ data provided by RIKEN, JAXA and the MAXI team, \Swiftbat\ transient monitor results provided by the \Swiftbat\ team, and the \Swiftxrt\ Data Analysis Software (XRTDAS) developed under the responsibility of the ASI Science Data Center (ASDC), Italy.
We acknowledge extensive use of the ADS and arXiv.

{\it Facilities:} \facility{CXO (ACIS)}, \facility{Swift (XRT,BAT)}, \facility{MAXI (GSC)}

\bibliographystyle{apj}
\bibliography{odd_src_ref_list}

\end{document}